\documentclass{elsart}
\usepackage{graphicx}

\begin{document}
\begin{frontmatter}
\title{Implications to Sources of Ultra-high-energy Cosmic Rays from their Arrival Distribution}

\author[utap]{Hajime Takami}\footnote{E-mail addresses: takami@utap.phys.s.u-tokyo.ac.jp (H.Takami), sato@phys.s.u-tokyo.ac.jp (K.Sato)} and 
\author[utap,resceu,ipmu]{Katsuhiko Sato}

\address[utap]{Department of Physics, School of Science, the University of Tokyo, 7-3-1 Hongo, Bunkyo-ku, Tokyo, 113-0033, Japan}
\address[resceu]{Research Center for the Early Universe, School of Science, the University of Tokyo, 7-3-1 Hongo, Bunkyo-ku, Tokyo, 113-0033, Japan}
\address[ipmu]{Institute of Physics and Mathematics for Universe, the University of Tokyo, Kashiwa, Chiba, 277-8582, Japan}

\begin{abstract}
We estimate the local number density 
of sources of ultra-high-energy cosmic rays (UHECRs) 
based on the statistical features of their arrival direction distribution. 
We calculate the arrival distributions of protons above $10^{19}$ eV 
taking into account their propagation process 
in the Galactic magnetic field and a structured intergalactic magnetic field, 
and statistically compare those with the observational result 
of the Pierre Auger Observatory. 
The anisotropy in the arrival distribution at the highest energies 
enables us to estimate the number density of UHECR sources 
as $\sim 10^{-4}~{\rm Mpc}^{-3}$ 
assuming the persistent activity of UHECR sources. 
We compare the estimated number density of UHECR sources 
with the number densities of known astrophysical objects. 
This estimated number density is consistent 
with the number density of Fanaroff-Reily I galaxies. 
We also discuss the reproducability of the observed {\it isotropy} in 
the arrival distribution above $10^{19}$ eV. 
We find that the estimated source model cannot reproduce 
the observed isotropy. 
However, the observed isotropy can be reproduced 
with the number density of $10^{-2}$-$10^{-3}~{\rm Mpc}^{-3}$. 
This fact indicates the existence of UHECR sources 
with a maximum acceleration energy of $\sim 10^{19}$ eV 
whose number density is an order of magnitude more than 
that injecting the highest energy cosmic rays. 
\end{abstract}

\begin{keyword}
ultra-high-energy cosmic rays
\end{keyword}

\end{frontmatter}

\section{Introduction} \label{introduction}

The origin of ultra-high-energy cosmic rays (UHECRs) has been highly unknown 
in spite of prolonged effort to construct larger UHECR observatories 
and to detect more events \cite{nagano00}. 
In cosmic ray spectrum, 
a sharply spectral steepening at around $10^{20}$eV 
has been predicted theoretically 
by interactions with photopion production 
with cosmic microwave background (CMB) photons, 
known as Greisen-Zatsepin-Kuz'min (GZK) steepening 
\cite{greisen66,zatsepin66}. 
The feature of this steepening 
observed by the High Resolution Fly's Eye (HiRes) \cite{abbasi04a,abbasi08} 
and the Pierre Auger Observatory (PAO) \cite{yamamoto07,abraham08} 
implies that astrophysical sources are much more dominant than 
top-down sources (for a review, see Ref. \cite{bhattacharjee00}) 
at the highest energies, 
though physical reasons for the extension of the energy spectrum 
beyond the GZK energies reported by the Akeno Giant Air Shower Array (AGASA) 
have been not understood yet \cite{takeda98,takeda03}. 
Several objects to accelerate particles up to $10^{20}$ eV 
have been suggested, 
but there has been little observational evidence 
which is a UHECR source 
(see Ref. \cite{bhattacharjee00,torres04} and reference therein).

The arrival distribution of UHECRs has information 
on the distribution of their sources, 
also including that on intergalactic and the Galactic magnetic field 
(IGMF and GMF).  
The AGASA and PAO reported 
statistically significant anisotropy at small angular scale 
in the arrival distribution \cite{takeda99,pao07,pao08}. 
The small-scale anisotropy implies point-like sources, 
and enables us to constrain the number of nearby UHECR sources. 
The source number density, $n_s$, is one of important parameters 
to investigate the nature of UHECR sources. 
Comparing the estimated number density 
with the number densities of known astrophysical objects, 
we can constrain the object-type of UHECR sources. 
Several authors have estimated as $n_s \sim 10^{-5}$-$10^{-6}~{\rm Mpc}^{-3}$ 
using the published AGASA data above $4 \times 10^{19}$ eV 
assuming the same injection rate over all sources 
\cite{yoshiguchi03,blasi04,kachelriess05,takami06,takami07}, 
and we also constrained the source number density 
as $10^{-4}~{\rm Mpc}^{-3}$, 
assuming a model that the injection rate is proportional to 
the luminosity of galaxies \cite{takami07}.

The PAO reported the correlation 
between the arrival directions of UHECR above $5.7 \times 10^{19}$ eV 
and the positions of nearby active galactic nuclei (AGNs) 
listed in the 12th Veron-Cetty \& Veron catalog \cite{veron06} 
within the angular scale of $3.1^{\circ}$ \cite{pao07,pao08}. 
However, most of the PAO-correlated AGNs are Seyfert galaxies and LINERs, 
which have much weaker activity than radio-loud AGNs 
like Fanaroff-Reily II (FR II) galaxies \cite{moskalenko08}. 
To understand whether these galaxies with weak activity 
are really UHECR sources or not, 
another information on UHECR sources is required.

The PAO estimated a lower limit of the UHECR source number as 61 
based on anisotropy in the arrival distribution of their detected events, 
simply assuming Poisson statistics and 
not taking into account UHECR propagation \cite{pao08}. 
This is certainly a lower limit of the number of the sources,  
but is not quantity which can be compared with astronomical observables 
because we do not know what radius they are included in. 
Their number density is an observable. 
Taking into account UHECR propagation, 
we can also estimate plausible value of it, not {\it limit}.

In this study, 
we simulate the arrival distribution of UHECRs above $10^{19}$ eV, 
taking into account their propagation process 
in intergalactic and the Galactic space. 
We extract an anisotropy signal from the simulated arrival distribution 
and then compare this signal with that observed by the PAO 
above $5.7 \times 10^{19}$ eV 
to estimate the source number density. 
The reports for the correlation by the PAO 
with nearby large-scale structure \cite{pao07,pao08,kashti08,takami08c} show 
that it plays a crucial role 
on the arrival distribution of UHECRs. 
Thus, we adopt the models of the IGMF and source distribution 
which reproduce the local structure actually observed 
around the Milky Way, developed in our previous work \cite{takami06}. 
We also discuss the isotropy in the arrival distribution 
at $\sim 10^{19}$ eV and estimate the source number density 
for the lower energies. 
The composition is assumed to be pure protons.

This paper is structured as follows. 
In section \ref{methods}, our calculation method 
for calculating UHECR arrival distribution 
and a statistical method to estimate the source number density are explained. 
In section \ref{results}, the calculation results are shown 
and we discuss the results and conclude in section \ref{discussion}.

\section{Methods} \label{methods}

We estimate the number density of UHECR sources as follows. 
First, we calculate arrival distributions of UHE protons 
based on our source models with several number densities. 
Their propagation process in intergalactic and the Galactic space 
is taken into account. 
The number of simulated protons and the threshold of their energies 
are set to the same as those detected by the PAO. 
Next, auto-correlation functions, 
which are an indicator of small-scale anisotropy in the arrival distribution, 
are calculated from our simulated events. 
Finally, comparing these auto-correlation functions 
with that calculated from the observed events, 
we find the best-fit value of the source number density.

We calculate the arrival distribution of UHE protons 
by a method used in Ref. \cite{takami07}. 
A characteristic of this method is to adopt a structured IGMF model 
which reproduces the observed structure in local universe, 
developed in Ref. \cite{takami06}. 
The strength of the IGMF is normalized 
at the center of Virgo cluster 
as $B=0.0, 0.1, 0.4$ and $1.0\mu$G. 
This method also enables us to take account of the deflection by GMF. 
A GMF model with bisymmetric spiral structure proposed by Ref. \cite{sofue83} 
and the same model parameters as Ref. \cite{stanev97} is adopted. 
The maximum distance of UHECR sources is set to 1Gpc, 
which is comparable with the energy-loss length of 
Bethe-Heitler pair creation with the CMB. 
This is sufficient to consider cosmic rays down to $10^{19}$ eV. 
There are two improvements for this study. 
One is the angular resolution of UHECR experiments. 
In Ref. \cite{takami07}, 
since we consider the AGASA data, 
we adopt the angular resolution of $\sim 2^{\circ}$. 
In this study, $1^{\circ}$ is adopted for the PAO events, 
which corresponds to the angular resolution of the PAO \cite{pao08}. 
The other is to take the non-uniformity of the exposure of ground array 
into account.

The exposure of ground array is not uniform 
because of the daily rotation of the earth. 
Since the variation in right ascension in a day can be neglected \cite{pao08}, 
the dependence of the exposure is simply written 
as a function of the declination of arriving cosmic rays, 
$\delta$, \cite{sommers01}, 
\begin{equation}
\omega (\delta) \propto \cos (a_0) \cos (\delta) \sin (\alpha_m) 
+ \alpha_m \sin (a_0) \sin (\delta)
\end{equation}
where $\alpha_m$ is given by 
\begin{eqnarray}
\alpha_m = \left\{ 
\begin{array}{cl}
0 & {\rm if}~~\xi > 1 \\
\pi & {\rm if}~~\xi <-1 \\
\cos^{-1} (\xi) & {\rm otherwise}, 
\end{array}
\right.
\end{eqnarray}
and 
\begin{equation}
\xi \equiv \frac{\cos (\theta_m) - \sin (a_0) \sin (\delta)}
{\cos (a_0) \cos (\delta)}. 
\end{equation}
Here, $a_0$ is detector's terrestrial latitude, 
and $\theta_m$ is the maximum zenith angle 
for the experimental cut. 
For the PAO, 
$a_0=35.2^{\circ}$ and $\theta=60^{\circ}$ are adopted \cite{sommers01}.

In this study, we adopt two source models. 
Both models are constructed from 
{\it Infrared Astronomical Satellite Point Source Redshift Survey} 
({\it IRAS} PSCz) catalog of galaxies \cite{saunders00}. 
The IRAS catalog consists of 14,677 galaxies above 0.6 Jy with redshift 
and it sky coverage is 84\%. 
So it is the best catalog to reflect the local large-scale structure. 
We correct the selection effect and galaxies in 16\% of unobserved area 
using the luminosity function estimated by Ref. \cite{takeuchi03}, 
and then regard subsets of the corrected IRAS galaxies 
as UHECR source distributions. 
Each galaxy is adopted as a source 
with the probability proportional to its luminosity. 
We perform the source selection 100 times for every $n_s$. 
One of the models is a source model with the same cosmic ray injection rate 
over all sources (called {\it model A} below), 
and the other is with the injection rate 
proportional to infrared luminosities of galaxies 
(called {\it model B} below). 
The maximum acceleration energy of protons is not dependent on 
luminosities of the sources. 
This corrected catalog is also used for the construction of our IGMF model.

Auto-correlation function is one of good indicators 
of small-scale anisotropy of UHECR arrival distribution. 
Auto-correlation function is defined as 
\begin{equation}
w(\theta) = \frac{1}
{2\pi \left| \cos \theta - \cos (\theta + \Delta \theta) \right|} 
\sum_{\theta \leq \phi \leq \theta+\Delta \theta} 1 [{\rm sr}^{-1}], 
\end{equation}
where $\phi$ is the separation angle of a pair of events. 
The interval $\Delta \theta$ is set to be $1^{\circ}$. 
In order to investigate the goodness of fit 
between the auto-correlation functions calculated from simulated events 
and observed one, 
we also define $\chi_{\theta_{\rm max}}$ as 
\begin{equation}
\chi_{\theta_{\rm max}} = \frac{1}{N_{\rm bin}} 
\sqrt{\sum_{n=0}^{N_{\rm bin}-1}
\frac{\left[ w(\theta_n) - w_{\rm obs}(\theta_n) \right]^2}
{{\sigma(\theta_n)}^2}}, 
\label{eq:chi}
\end{equation}
where $w(\theta)$ is an average of the auto-correlation function 
calculated from a source distribution 
and $\sigma(\theta)$ is the standard deviation of $w(\theta)$ 
due to the finite number of simulated events. 
Random event selection is performed 100 times for every source distribution. 
$w_{\rm obs}(\theta)$ is the auto-correlation function 
calculated from observed data. 
$\theta_{\rm max}$ is the maximum angular scale 
to consider the small-scale anisotropy, 
$N_{\rm bin} = \theta_{\rm max} / \Delta \theta$ is the number of angular bins.

\section{Results} \label{results}

\subsection{Estimation of $n_s$ from the small-scale anisotropy} \label{ssa}

\begin{figure}[t]
\begin{center}
\rotatebox{-90}{\includegraphics[width=0.45\linewidth]{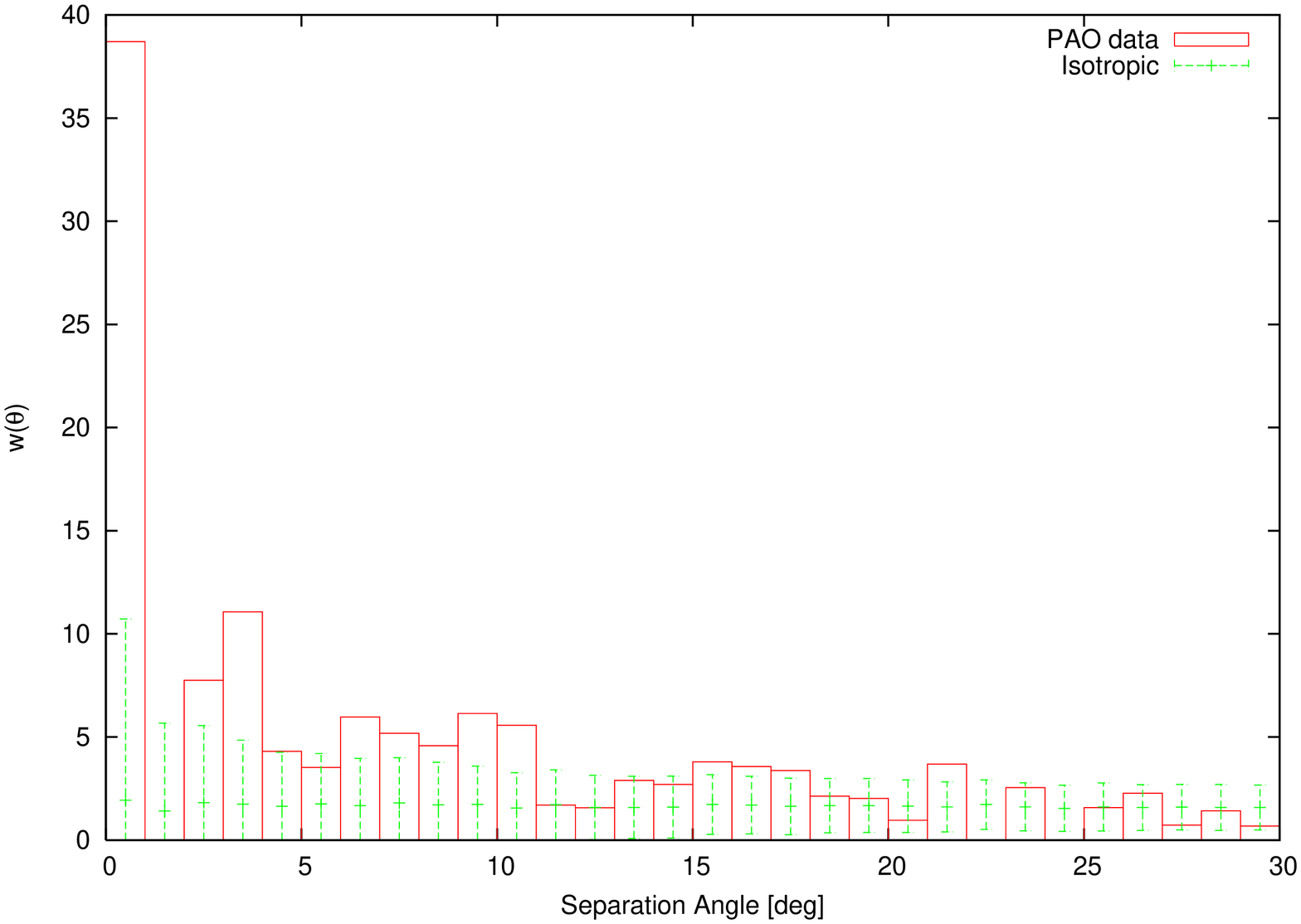}}
\caption{Auto-correlation functions calculated 
from the 27 PAO events ({\it histogram}) and random distribution 
weighted by the exposure of the PAO. 
The error bars are the standard deviations due to the finite number of events. 
The event realization is performed 1000 times. }
\label{fig:acor}
\end{center}
\end{figure}

We estimate UHECR source number density 
using an anisotropy signal in the 27 events published by the PAO 
whose energies are above $5.7 \times 10^{19}$ eV \cite{pao08}. 
First, we check the anisotropy signal. 
Fig. \ref{fig:acor} shows auto-correlation function calculated 
from the 27 events ({\it histogram}). 
The auto-correlation function predicted from random distribution 
weighted by the exposure of the PAO is also shown. 
The error bars are the standard deviations due to the finite number of events. 
The event realization is performed 1000 times. 
We can see statistically significant anisotropy at small angular scale 
against isotropic distribution. 
We set $\theta_{\rm max}=5^{\circ}$ in Eq. \ref{eq:chi}  
to extract the small-scale anisotropy. 
If $\theta_{\rm max} \sim 10^{\circ}$ is adopted, 
main results below are unchanged.

Next, we calculate auto-correlation functions 
from our simulated arrival distributions of UHE protons, 
which are compared with the histogram in Fig. \ref{fig:acor} 
to investigate the number density of the sources 
which can best reproduce the small-scale anisotropy observed by the PAO. 
If $n_s$ is extremely small, 
strong anisotropy is expected 
because only few sources can contribute to the arriving cosmic rays. 
On the other hand, in the case of much many sources, 
the arrival distribution is expected to be close to isotropy. 
Thus, there should be the best-fit value of the number density.

\begin{figure}[t]
\begin{center}
\includegraphics[width=0.48\linewidth]{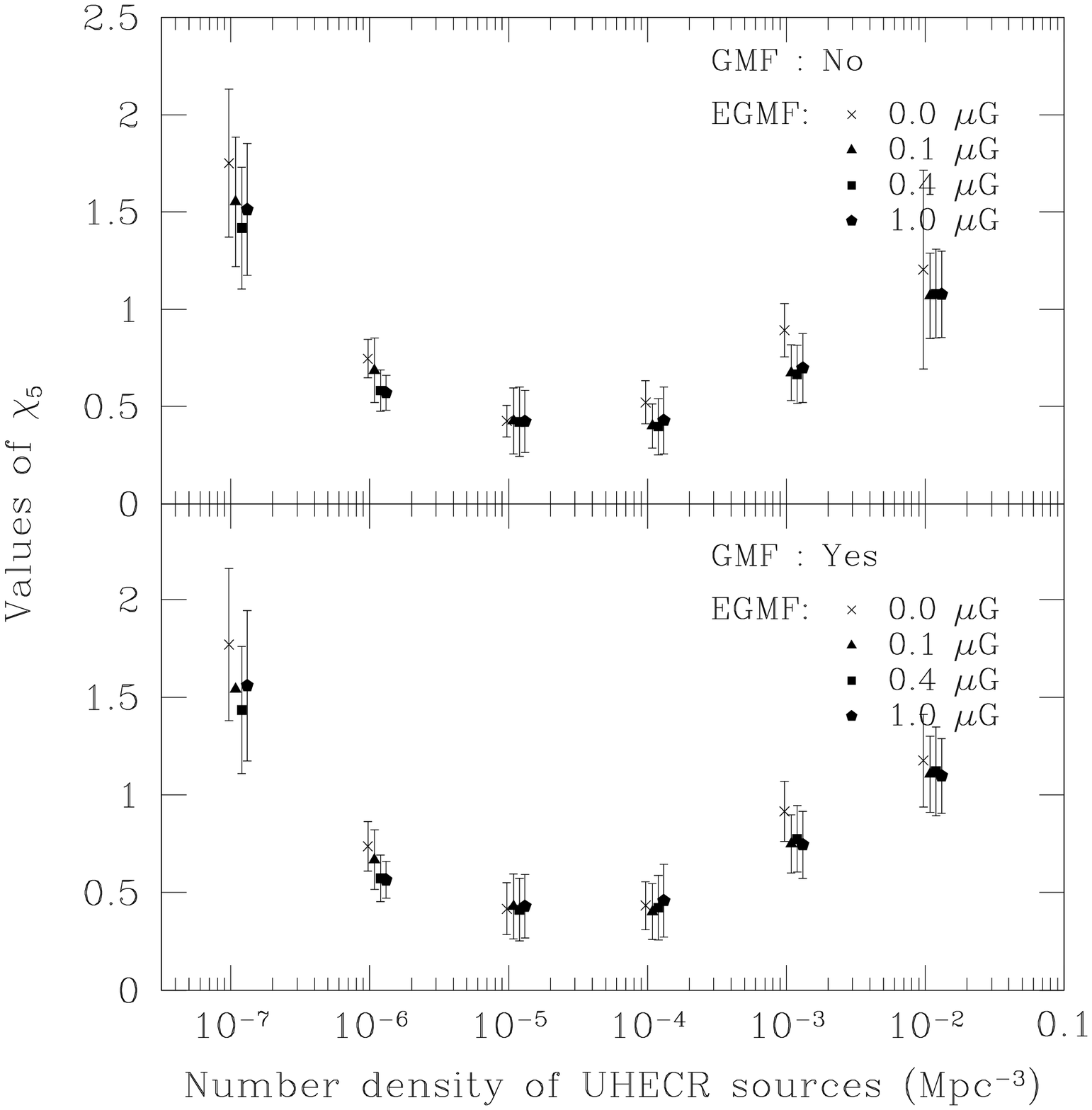} \hfill
\includegraphics[width=0.48\linewidth]{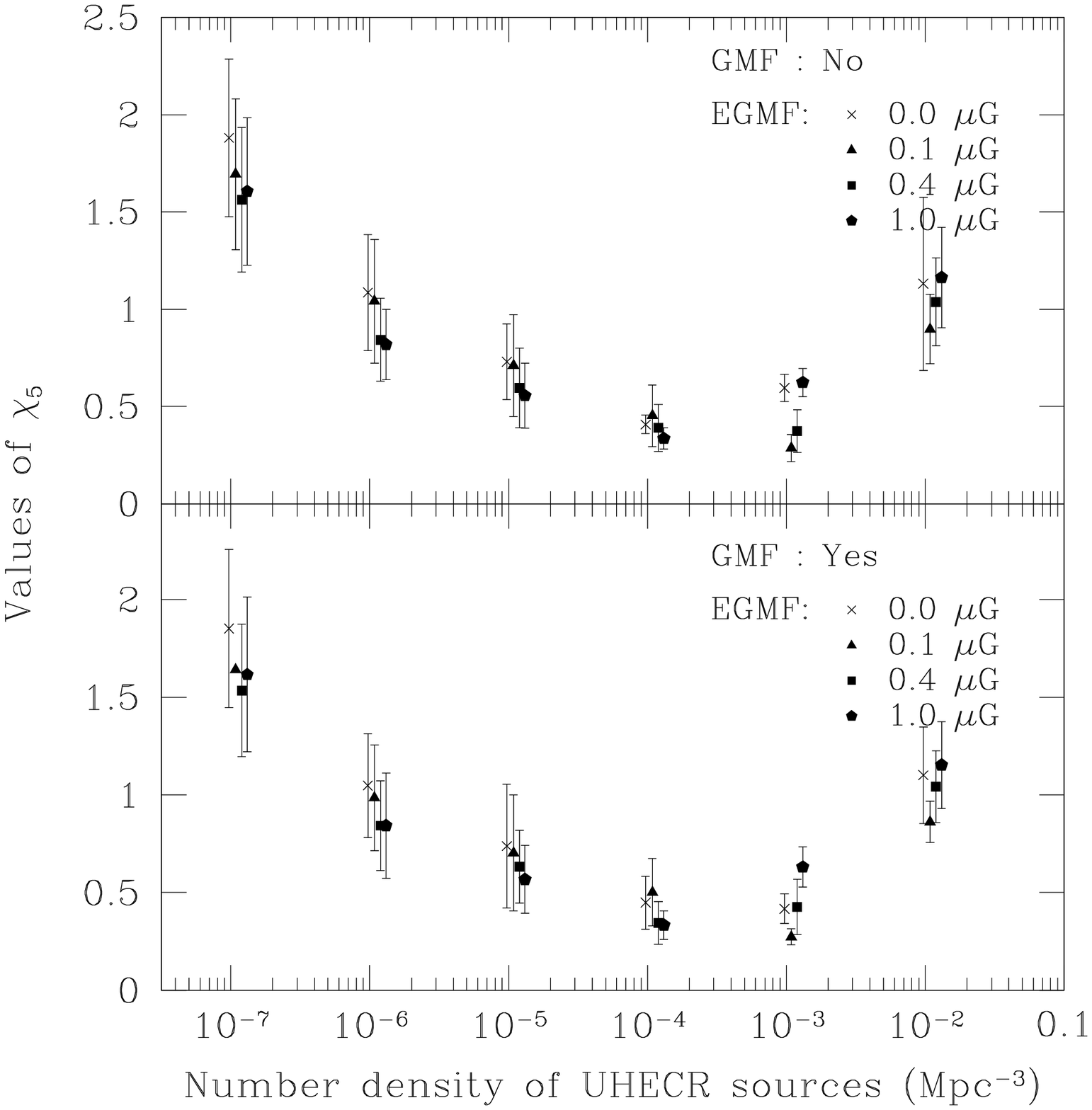}
\caption{$\chi_5$ for different number densities of UHECR sources 
in the model A ({\it left}) 
and model B ({\it right}). 
The error bars originate from 100 times source selection. 
The GMF is considered in the lower panel 
while not in the upper panel.}
\label{fig:chi}
\end{center}
\end{figure}

Fig. \ref{fig:chi} shows $\chi_5$ calculated 
for the different number densities of UHECR sources. 
The normalization of the IGMF strength considered is 0.0 ({\it crosses}), 
0.1 ({\it triangles}), 0.4 ({\it squares}), 
and 1.0 $\mu$G ({\it pentagons}) respectively. 
The error bars represents the standard deviation of $\chi_5$, 
which is estimated from 100 realizations of the source distribution. 
The plots at the same number density 
are a little shifted horizontally for visibility. 
Note that UHECR sources within 5 Mpc are artificially neglected 
in this calculation 
to reproduce isotropy observed at lower energies ($\sim 10^{19}$ eV). 
This point will be discussed in section \ref{lsi}. 
Such nearby sources also predict anisotropy 
stronger than observed on at the highest energies.

In the left figure, $\chi_5$ calculated in the model A is shown. 
The GMF is considered in the lower panel while not in the upper panel. 
For $n_s \sim 10^{-7}~{\rm Mpc}^{-3}$, 
there is zero or one source within 100 Mpc. 
This extremely few number of sources leads to 
anisotropy much stronger than the observational result. 
Thus, $n_s \sim 10^{-7}~{\rm Mpc}^{-3}$ is too small 
to reproduce the observed arrival distribution. 
If, on the other hand, $n_s \sim 10^{-2}~{\rm Mpc}^{-3}$, 
which is comparable with the number of bright galaxies, 
high-precision isotropy of the arrival distribution is realized. 
This another extreme case cannot also reproduce the observed anisotropy. 
Thus, $\chi_5$ as a function of $n_s$ has the minimum 
at the intermediate number density. 
In the upper panel, 
$\chi_5$ is minimized at around $n_s \sim 10^{-5}$-$10^{-4}~{\rm Mpc}^{-3}$, 
almost independent of the IGMF strength in our IGMF model. 
In our IGMF model, 
since about 95\% of volume within 100 Mpc has no magnetic field, 
the highest energy protons are deflected only 
in the neighbourhood of their sources where the universe is magnetized. 
If the universe is strongly magnetized, for example, 
uniformly with $\sim 100$nG, 
the source number density is allowed to be smaller 
because strong deflections makes the arrival distribution to be isotropic 
even at the highest energies. 
In the lower panel, 
$\chi_5$ is minimized at the same number densities. 
The constrained number density is not dependent on the existence 
of the GMF since the clustering signal is almost not varied 
because of the coherence of the GMF. 
Note that the GMF can changes the arrival directions of UHECRs efficiently 
and positional correlation with their sources 
as shown in Ref. \cite{takami08b}.

We also estimate the source number density in the model B 
by similar discussion above. 
In the right figure, $\chi_5$ in the model B is shown. 
$\chi_5$ is minimized at around $n_s \sim 10^{-4}$-$10^{-3}~{\rm Mpc}^{-3}$, 
which is about an order of magnitude more than in the model A. 
In the model B, 
luminous sources strongly contribute to the arriving cosmic rays 
and, on the other hand, weaker sources do not almost contribute 
the flux of cosmic rays 
though these are counted as sources. 
Thus, the constrained number density is effectively larger.

The number densities estimated from the anisotropy signal of the PAO 
based on the two source models (A \& B) are consistent with 
the results of our previous study using the AGASA data \cite{takami07}. 
The uncertainty of about one order of magnitude 
will be reduced by 5 years observation by the PAO, 
as pointed out in Ref. \cite{takami07}.

The model A and B are two extreme cases. 
In the model A, all sources are identical. 
This model can be well applied to astrophysical objects 
which have hardly personality regarding emitted energy, 
or strongly active sources like FR II galaxies 
and Gamma Ray Bursts (GRBs) (though GRBs are transient sources). 
On the other hand, the model B is applied to objects 
common in the universe like bright galaxies. 
When the model B was constructed, 
the IRAS galaxies with luminosities 
of $10^{7}$-$10^{12} L_{\odot}$ were adopted, 
where $L_{\odot}$ is the Solar luminosity, $3.826 \times 10^{26}$ W. 
The luminosities are over 5 orders of magnitude. 
The width of magnitude is realized by less active objects 
like bright galaxies and Seyfert galaxies. 
In fact, UHECR sources have intermediate nature between the two models. 
Assumed to be highly active objects, 
UHECR sources has the number density close to that in the model A. 
Thus, the number density of UHECR sources can be 
estimated as $\sim 10^{-4}~{\rm Mpc}^{-3}$.

Compared with the number densities of several candidates of UHECR sources, 
this estimated value can constrain the object-type of UHECR sources. 
Table \ref{tab:candidates} shows local number densities 
of several astrophysical objects. 
The objects which have smaller number density than the estimated one are not 
main contributors of the arrival highest energy cosmic rays. 
Both FR II galaxies and BL Lac objects have been 
plausible UHECR sources theoretically \cite{rachen93,torres04}, 
but they are not mainly contribute to the flux of UHECRs. 
Note that this constraint does not reject them as UHECR sources. 
They are not main contributors. 
The number density of bright galaxies 
(defined as $-22 < M_{B_T} < -18$, 
where $M_{B_T}$ is absolute magnitude in B-band \cite{ulvestad01}) 
is much larger than our constraint. 
The generation of the highest energy cosmic rays is 
not common phenomena in the local universe. 
The number density of Seyfert galaxies is an order of magnitude 
larger than the constrained value, 
and therefore it is allowed if $\sim 10$\% of such Seyfert galaxies has 
activity to accelerate UHECRs. 
However, Seyfert galaxies are generally not expected to be UHECR sources. 
For example, the PAO-correlated AGNs, most of all is Seyfert or LINER, 
do not show significant jet activity \cite{moskalenko08}. 
Therefore, there are no reasons for expecting them to
accelerate cosmic rays up to the highest energies at all 
in the jet paradigm. 
The number density of FR I galaxies is 
$\sim 10^{-4}~{\rm Mpc}^{-3}$ \cite{padovani90}, consistent with the constraint. 
FR I galaxy is one of plausible acceleration site 
up to the highest energies. 
Centaurus A (Cen A), the nearest PAO-correlated AGNs, 
is classified into FR I galaxy. 
At a hot spot in the jet in this galaxy, 
an estimation of the maximum acceleration energy of protons 
is $\sim 10^{20.6}$ eV \cite{torres04}. 
If the other FR I galaxies have similar configurations, 
it is possible that FR I galaxies are main contributors 
of the highest energy cosmic rays.

\subsection{Isotropy at around $10^{19}$ eV} \label{lsi}

At the previous section, 
we artificially neglect sources within 5 Mpc from the Galaxy. 
This is because very nearby sources inevitably produce anisotropy 
in arrival distribution of UHECRs at around $10^{19}$eV. 
Since all observatories on UHECRs have reported isotropy 
in that energy range \cite{takeda99,abbasi04b,mollerach08}. 
it is inconsistent with the observational results. 
We discuss the isotropy in this subsection.

\begin{figure}[t]
\begin{center}
\includegraphics[width=0.95\linewidth]{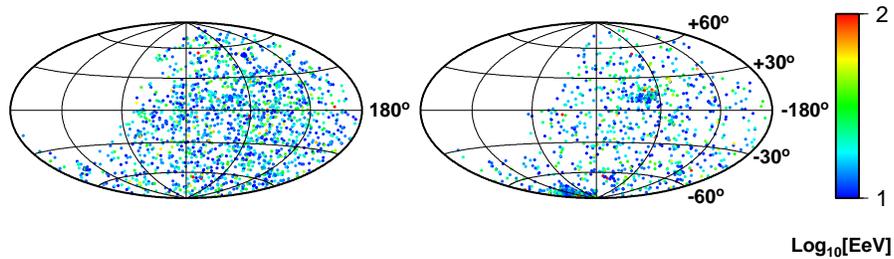}
\caption{The arrival distributions of 1672 protons above $10^{19}$ eV 
(= 10 EeV) calculated from a source distribution in the model A 
with $n_s \sim 10^{-3}~{\rm Mpc}^{-3}$ 
in which the sources within 5 Mpc are included ({\it right}) 
and neglected ({\it left}). 
The exposure of the PAO is taken into account. 
The IGMF strength is $B=0.4 \mu$G and the GMF is neglected.}
\label{fig:adN1672F50NO1GMF0EGMF4L0}
\end{center}
\end{figure}

Fig. \ref{fig:adN1672F50NO1GMF0EGMF4L0} shows the arrival distributions 
of protons above $10^{19}$ eV 
simulated from a source distribution in the model A 
with $n_s \sim 10^{-3}~{\rm Mpc}^{-3}$, 
which is about one order of magnitude larger than 
the number density constrained in the previous section. 
In the left figure, the sources within 5 Mpc are neglected. 
The number of events is set to 1672, 
which corresponds to that detected by the PAO \cite{mollerach08}. 
The IGMF strength is $B=0.4 \mu$G.

\begin{figure}[t]
\begin{center}
\includegraphics[width=0.95\linewidth]{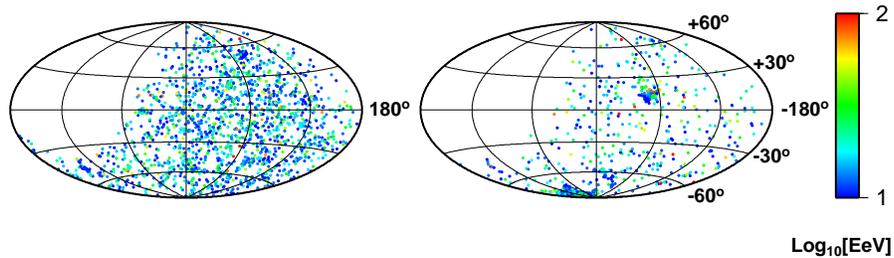}
\caption{The same as Fig. \ref{fig:adN1672F50NO1GMF0EGMF4L0}, 
but in the model B.}
\label{fig:adN1672F50NO1GMF0EGMF4L1}
\end{center}
\end{figure}

In the right figure, 
strong anisotropies appear 
at the positions of $(\ell, b) \sim (-50^{\circ}, 20^{\circ})$ and 
$(100^{\circ}, -85^{\circ})$, 
which are generated by protons injected from Cen A 
and NGC 253, a famous starburst galaxy, respectively. 
These distances are 4.6 Mpc and 4.2 Mpc in the IRAS PSCz catalog, respectively 
\cite{saunders00}. 
Note that the distance of Cen A is different among many catalogs. 
If these nearby objects are really UHECR sources, 
strong anisotropy is inevitably predicted. 
The PAO reported that the arrival distribution of UHECRs 
above $10^{19}$ eV is consistent with isotropic distribution
with 95\% confidence level \cite{mollerach08}.  
Thus, the anisotropy signals are inconsistent with the PAO results. 
Such anisotropy is predicted as long as there are UHECR sources 
in the nearby universe even if $n_s \sim 10^{-3}~{\rm Mpc}^{-3}$ is adopted. 
On the other hand, 
the anisotropies disappears in the left figure. 
This fact is also true in the model B, 
as shown in Fig. \ref{fig:adN1672F50NO1GMF0EGMF4L1}. 
Thus, these nearby objects are not UHECR sources.

In fact, the observed high-level isotropy cannot be easily reproduced 
by neglecting sources within 5 Mpc. 
In order to see this, 
we compare the auto-correlation function of simulated events 
above $10^{19}$ eV with that predicted from isotropic distribution. 
Table \ref{tab:fraction} shows the fraction of the number 
of source distributions which can reproduce the arrival distribution 
consistent with the isotropy within some errors written in the table. 
$\sigma$ means the standard deviation of the auto-correlation function 
calculated from random distribution, 
which is due to the finite number of events. 
The normalization of the IGMF strength considered is 0.1 and 0.4 $\mu$G. 
In the model A, 
there are few source distributions with $n_s \sim 10^{-4}~{\rm Mpc}^{-3}$ 
which can reproduce the isotropy within 2$\sigma$. 
Even for $n_s \sim 10^{-3}~{\rm Mpc}^{-3}$, 
it is less than 10 though isotropy is found by eye 
in Figs. \ref{fig:adN1672F50NO1GMF0EGMF4L0} 
and \ref{fig:adN1672F50NO1GMF0EGMF4L1}. 
If the number density is $\sim 10^{-2}~{\rm Mpc}^{-3}$, 
the isotropy can be well reproduced.

In addition to the structured IGMF, 
we take account of a uniform turbulent magnetic field 
with the strength and coherent length of $B_{\rm tur}$, $l_c$, respectively. 
Faraday rotation measurements of distant quasars 
shows $B {l_c}^{1/2} < (1~{\rm nG})(1~{\rm Mpc})^{1/2}$ 
\cite{kronberg94}. 
According to this constraint, we adopt  $B_{\rm tur}=1$ nG and $l_c=1$ Mpc. 
The results are also shown in Table \ref{tab:fraction}. 
We are able to find that the number of the source distributions 
which can reproduce the isotropy is increased, 
compared with that without the turbulent field. 
However, for the estimated number density, 
the number is still too small to naturally reproduce the isotropy. 
When the source number density an order of magnitude larger than 
the estimated one, 
the number is dramatically increased to 70-80\% 
within 3$\sigma$ in the model A. 
If the number density is $\sim 10^{-2}~{\rm Mpc}^{-3}$, 
the isotropy can be well reproduced.

It is much larger than the number density estimated 
based on the small-scale anisotropy at the highest energies. 
This fact indicates the existence of sources 
which contributes to lower energy cosmic rays, 
whose number density is $10^{-2}$-$10^{-3}~{\rm Mpc}^{-3}$. 
In other words, 
there are many sources with the maximum acceleration energy 
of $\sim 10^{19}$ eV. 
Note that the uniform IGMF does almost not affects 
the estimation of the number density at the highest energy.

\section{Discussion \& Conclusion} \label{discussion}

In this study, 
we estimated the number density of UHECR sources 
based on the statistical features of the arrival distribution of UHECRs 
observed by the PAO. 
We simulated the arrival distributions of protons above $10^{19}$ eV, 
taking into consideration their propagation process 
in the Galactic and intergalactic space. 
The IGMF model adopted was associated with the observed large-scale structure. 
Comparing the simulated arrival distributions with the data 
observed by the PAO statistically, 
we estimated the number density which can best reproduce 
the observational result. 
The anisotropy signal above $5.7 \times 10^{19}$ eV led to 
$n_s \sim 10^{-4}~{\rm Mpc}^{-3}$ 
which is consistent with that of FR I galaxies, 
which are a candidate to accelerate protons up to $10^{20}$ eV. 
We also focused on isotropy in the arrival distribution 
at around $10^{19}$ eV, 
and then found $n_s \sim 10^{-2}$-$10^{-3}~{\rm Mpc}^{-3}$, 
which is one or two order of magnitude larger than 
that estimated from the anisotropy.

In this calculation, 
cosmological evolution of UHECR sources is not considered. 
Since protons with $\sim 10^{19}$ eV can reach the earth 
from sources at the distance of 1Gpc ($z \sim 0.25$), 
which is comparable with the energy-loss length of such protons 
by Bethe-Heitler pair creation, 
it might be possible that the cosmological evolution could 
reproduce the isotropy even if the local number density of UHECR sources is 
comparable with $\sim 10^{-4}~{\rm Mpc}^{-3}$. 
However, a typical evolution factor is $(1+z)^3$, 
and therefore the number density at $z \sim 0.25$ is only twice 
more than the local one. 
Such a small factor could not change the estimated number density 
by two order of magnitude less. 
The difference between the two estimated number densities is significant.

The difference can be interpreted as the evidence of the existence 
of UHECR sources which can accelerate protons up to $\sim 10^{19}$ eV 
assuming the persistent activity of UHECR sources. 
$n_s \sim 10^{-2}$-$10^{-3}~{\rm Mpc}^{-3}$ is comparable with 
the number densities of bright galaxies or Seyfert galaxies. 
The proton acceleration up to $10^{19}$ eV might be common in the universe. 
We could also interpret this difference as transient activity to emit UHECRs. 
If bursting sources, like GRBs, are assumed, 
apparent number density, which corresponds to $n_s$ estimated in this study, 
depends on the threshold of UHECR energies 
because the dispersion of the time-delay is larger at lower energies 
\cite{miralda96}. 
However, quantitative discussion on this possibility exceeds 
the scope of this study. 
It is next study of ours.

In order to reproduce isotropy in the arrival distribution 
of UHECRs above $10^{19}$ eV, 
sources within 5 Mpc were artificially neglected. 
In these sources, Cen A, the nearest radio-loud AGN, is involved. 
The isotropy at around $10^{19}$ eV implies that it is not UHECR sources 
in our persistent source model. 
However, radio-loud AGNs have been plausible site 
for particle acceleration up to the highest energy \cite{torres04}, 
and the two of the highest energy events of the PAO are 
correlated with the position of Cen A within $\sim 3^{\circ}$. 
Whether Cen A is really a UHECR source or not is a key 
to understand the mechanism to generate the highest energy cosmic rays.

If Cen A is not a UHECR source, we can interpret that 
the source of cosmic rays arriving in the direction of Centaurus 
is behind Cen A and more distant, 
so that the strong anisotropy is not generated. 
The 2 events towards Cen A are also positionally correlated with NGC 5090 
with $\sim 3^{\circ}$, a radio galaxy with the distance of $\sim 40$ Mpc. 
This might be a real source though the activity is weak 
like the other PAO-correlated AGNs. 
These events are also towards Centaurus cluster whose distance is 
about 40 Mpc. 
An idea to generate UHE particles 
is cluster accretion shock \cite{kang96,inoue05,inoue07}. 
However, this scenario can accelerate protons up to $\sim 10^{19}$ eV.

If Cen A is a UHECR source, we propose several possibilities 
not to generate the anisotropy. 
One is UHECR composition. 
In the case of heavy dominated composition at energies above $10^{19}$ eV, 
the trajectories of UHECRs are deflected $Z$ times more than those of protons, 
and the discussion on the isotropy in this paper is not applied. 
Several composition measurements imply 
the existence of some fraction of heavy elements 
for all large uncertainty on hadronic interaction models 
in extensive air shower \cite{unger07,glushkov07}. 
The other is that the UHECR production is transient. 
For a bursting source, 
the energy of cosmic rays observed at present 
is in narrow energy range 
since the time-delay of cosmic rays depends on their energies \cite{miralda96}. 
In this brief picture, the 2 events are arrived from Cen A, 
and the lower energies will come in the future. 
The detailed discussion on the two possibilities is our near future plan.

The estimated number densities have some uncertainty 
because of the small number of detected events. 
The uncertainty to determine the number density at the highest energies 
can be reduced by increasing observed events \cite{takami07} and 
significant anisotropy would be observed 
in the arrival distribution at $\sim 10^{19}$ eV in the near future 
\cite{waxman97,kashti08}. 
The dramatically increase of detected events in the near future 
will provide us more useful information on UHECR sources.

\subsubsection*{Acknowledgements:} 
The work of H.T. is supported by Grants-in-Aid from JSPS Fellows. 
The work of K.S. is supported by Grants-in-Aid for 
Scientific Research provided by the Ministry of Education, Science 
and Culture of Japan through Research Grants S19104006.

\begin{table}
\caption{Local number densities of several active objects.}
\label{tab:candidates}
\begin{tabular}{lcr} \hline \hline
Object & Density [${\rm Mpc}^{-3}$] & Reference \\ \hline
Bright galaxy & $1.3 \times 10^{-2}$ & Ref. \cite{ulvestad01} \\
Seyfert galaxy & $1.25 \times 10^{-3}$ & Ref. \cite{ulvestad01} \\
Bright quasar & $1.4 \times 10^{-6}$ & Ref. \cite{hewett93} \\
Fanaroff-Reily 1 & $8 \times 10^{-5}$ & Ref. \cite{padovani90} \\
Fanaroff-Reily 2 & $3 \times 10^{-8}$ & Ref. \cite{woltjer90} \\
BL Lac objects & $3 \times 10^{-7}$ & Ref. \cite{woltjer90} \\ \hline
\end{tabular}
\end{table}

\begin{table}
\caption{The numbers of the source distributions 
which satisfy isotropy in the observed arrival distribution of UHECRs 
above $10^{19}$ eV within the error bars shown in table, 
in the case of the model A (without parentheses) and B (in parentheses). 
The total number of source distributions in each parameter set is 100.}
\label{tab:fraction}
\begin{tabular}{|c||cc|cc||cc|cc|} \hline \hline
$B_{\rm IGMF}$ & \multicolumn{4}{|c||}{0.1 $\mu$G} & \multicolumn{4}{|c|}{0.4 $\mu$G} \\ \hline
$B_{\rm tur}$ & \multicolumn{2}{|c|}{0.0 nG} & \multicolumn{2}{|c||}{1.0 nG} & \multicolumn{2}{|c|}{0.0 nG} & \multicolumn{2}{|c|}{1.0 nG} \\ \hline
$n_s$ [${\rm Mpc}^{-3}$] & $2\sigma$ & $3\sigma$ & $2\sigma$ & $3\sigma$ & $2\sigma$ & $3\sigma$ & $2\sigma$ & $3\sigma$ \\ \hline
$10^{-2}$ & 100(100) & 100(100)& 100(100) & 100(100) & 100(100) & 100(100) & 100(100) & 100(100) \\ \hline
$10^{-3}$ & 5 (0) & 54 (0) & 33 (0) & 79 (8) & 7 (0) & 34 (6) & 34 (0) & 73 (17) \\ \hline
$10^{-4}$ & 3 (0) & 10 (0) & 11 (0) & 28 (0) & 2 (0) & 15 (0) & 7 (0) & 26 (0) \\ \hline
$10^{-5}$ & 0 (0) & 2 (0) & 0 (0) & 9 (0) & 0 (0) & 0 (0) & 0 (0) & 6 (0) \\ \hline
\end{tabular}
\end{table}

\end{document}